\documentclass[12pt]{article}

\usepackage{amsmath,amssymb,amsfonts,bm,slashed,graphics}
\usepackage[latin1]{inputenc}
\usepackage[english]{babel}
\usepackage{epsfig}
\usepackage{hyperref}
\usepackage{amsthm}
\usepackage{subfigure}
\usepackage{dsfont}
\usepackage{hyperref}
\usepackage{amsthm}
\usepackage{subfigure}
\usepackage{color}
\usepackage{multirow}
\usepackage{psfrag}
\usepackage{graphicx}
\usepackage{extarrows}
\usepackage{color}
\usepackage{multirow}
\usepackage{tikz}

\DeclareMathOperator{\Tr}{Tr}

\newcommand{\cT}{{\cal{T}}}

\newtheorem{definition}{Definition}
\newtheorem{theorem}{Theorem}
\newtheorem{proposition}{Proposition}

\newcommand{\bea}{\begin{eqnarray}}
\newcommand{\eea}{\end{eqnarray}}
\newcommand{\bee}{\begin{equation}}
\newcommand{\ee}{\end{equation}}

\def\C{{\mathbf C}}

\newcommand{{\J}}{{\mathbf J}}
\newcommand{{\bbR}}{{\mathbb R}}
\newcommand{{\bbJ}}{{\mathbb J}}
\newcommand{{\bbL}}{{\mathbb L}}
\newcommand{{\bbB}}{{\mathbb B}}
\newcommand{{\bbS}}{{\mathbb S}}
\newcommand{{\bbZ}}{{\mathbb Z}}
\newcommand{{\bbT}}{{\mathbb T}}

\newcommand{{\bbw}}{{\mathbb w}}

\newcommand{{\frw}}{{\mathfrak  w}}
\newcommand{{\frW}}{{\mathfrak  W}}

\newcommand{{\mcT}}{{\mathcal T}}

\newcommand{\cR}{{\cal R}}

\newcommand{\cS}{{\cal S}}

\newcommand{\cH}{{\cal H}}

\newcommand{\cF}{{\cal F}}

\newcommand{\bbone}{{\bf 1}}
\newcommand{\tr}{{\rm Tr}}
\newcommand{\beann}{\begin{eqnarray*}}
\newcommand{\eeann}{\end{eqnarray*}}  

\newcommand{\gF}{{\mathfrak{F}}}
\newcommand{\LV}{\cS}
\newcommand{\prf}{{\noindent {\bf Proof}\quad }}
\begin{document}
\title{Loop Vertex Representation for Cumulants,\\
Part I: Bounds on Free Energy with Sources}
\author{V. Rivasseau\\
Universit\'e Paris-Saclay, CNRS/IN2P3\\ IJCLab, 91405 Orsay, France}
\date{} 

\maketitle

\begin{abstract}  
 In this paper we study the cumulants
 for stable random matrix models with single trace interactions of arbitrarily 
 high even order.  We obtain explicit and convergent expansions for it and we prove that it is an analytic function inside a cardioid domain in the complex plane. We also prove their Borel-LeRoy summability at the origin of the  coupling constant. 
Our proof is uniform in the external variables. 
\end{abstract}

\noindent\textbf{keywords}
Random Matrix; Cumulants; Constructive Field Theory


\medskip\noindent
{Data Availability Statements}
All data are available within the article or supplemental information.

\section{Introduction}

Random matrix theory \cite{Mehta,Akemann} studies probability laws for matrices.
Application of random matrices to 2d quantum gravity \cite{matrix} relies on their associated \emph{combinatorial maps},
which depend on (at least) two parameters: a coupling constant $\lambda$ and the size of the matrix, $N$.
A \emph{formal} expansion in the parameter $\lambda$ yields generating functions for maps of arbitrary genus. The coupling constant $\lambda$ roughly measures the size of the map while the parameter $1/N$ turns out to measure the genus of the map \cite{thooft}. 

We are interested in the loop vertex representation  (LVR) \cite{rivasseau2018loop}.
This is a improvement of the loop vertex expansion  (LVE) \cite{rivasseau2007constructive}. This LVE was introduced itself as a tool 
for constructive field theory to deal with random matrix fields. 
A common main feature of the LVR and LVE is that it is written in terms of trees which are exponentially bounded.
It means that the outcome of the LVR-LVE is convergent and is \emph{the Borel-LeRoy sum in $\lambda$},
whereas the  usual perturbative quantum field theory \emph{diverges} at the point $\lambda=0$. 
 The essential components of LVE are the Hubbard-Stratonovich intermediate field
representation \cite{Hub,Str}, the replica method \cite{MPV} and the BKAR formula \cite{BK,AR1}.
The added ingredients of the LVR are combinatorial, based on the selective Gaussian integration \cite{rivasseau2018loop},
and the Fuss-Catalan numbers and their generating function \cite{FussCatalan}.
We think that the LVR has \emph{more power} than the LVE, since the LVR can treat more models, 
with higher polynomial interactions. 

\medskip
For a general exposition of constructive field theory, see
\cite{simon2015p,glimm2012quantum,rivasseau2014perturbative}; for an early application to the generating function
of connected Schwinger functions - which in this paper are denoted {\it cumulants} - see \cite{magnen2008constructive}; 
for the actual mechanism of replacing Feynman graphs, which are not exponentially bounded, by trees, see \cite{RiZh1};
for a review of the LVE, we suggest consulting \cite{GRS}; and for the LVE applied to cumulants, we refer to  \cite{GuKra}.
Together with \cite{KRS}-\cite{KRS1}  this is our main source of inspiration for this paper.

\medskip
\begin{figure}[!htb]\centering
\includegraphics[width=0.3\linewidth]{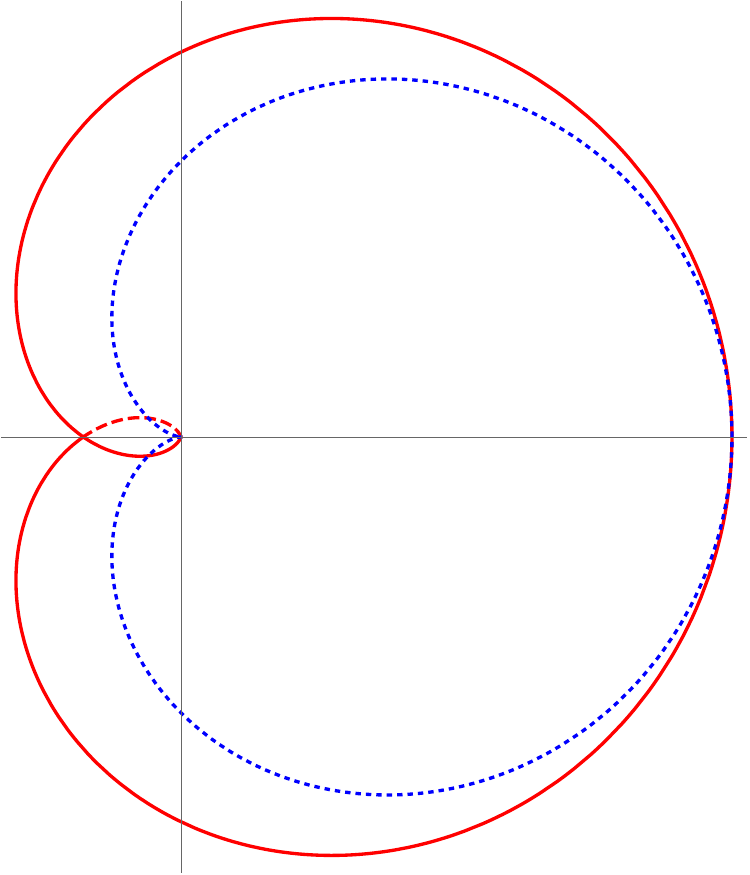} 
\caption{In blue the cardioid domain considered in \cite{rivasseau2022cumulants}, in red the cardioid domain considered
in \cite{BGKL}.}
\label{fig:W-domain}
\end{figure}

The authors of \cite{KRS}  join this formalism to Cauchy holomorphic matrix calculus
and have been applied to the simplest complex matrix model with stable monomial interaction. In \cite{KRS1} the  same authors 
have extended it to the case of \emph{Hermitian} or \emph{real symmetric} matrices,
in a manner both \emph{simpler and more powerful}. The basic formalism is still the LVR, but
while \cite{rivasseau2018loop,KRS} used contour integral parameters attached to every \emph{vertex} 
of the loop representation, \cite{KRS1} introduces more contour integrals, one for each \emph{loop vertex corner}.
This results in simpler bounds for the norm of the corner operators. 

\medskip
But we should remember that the LVE is older and their authors have more time 
to fine-tune their models. They construct their models with the coupling constant in a cardioid-shaped domain
(see Figure \ref{fig:W-domain}) which has opening angle arbitrarily close to $2 \pi$ \cite{rivasseau2022cumulants}  
or even exceeding $2 \pi$ \cite{BGKL}. In this case the LVE is capable to compute  
 some typically non-perturbative effects like instantons by resuming perturbative field theory. In \cite{sazonov2023}, Sazonov combined the LVE with ideas of the variational perturbation theory.

\begin{figure}[!htb]
\centering\includegraphics[width=0.5\linewidth]{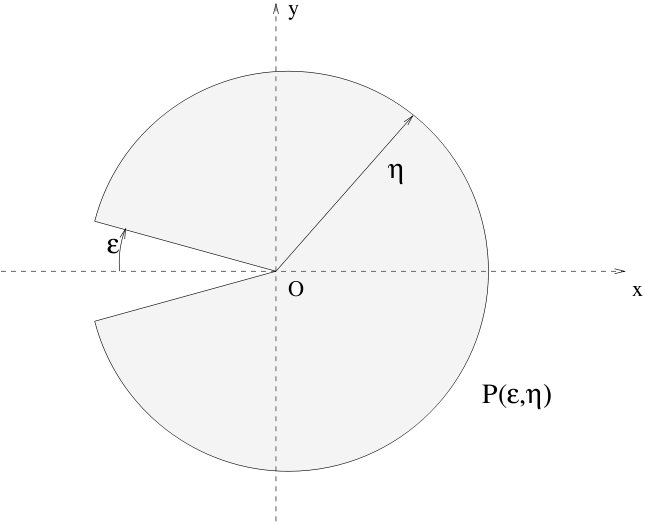} 
\caption{The pacman domain with parameters $(\eta, \epsilon)$, defined by $P(\epsilon, \eta) := \{0< |\lambda |<\eta, \vert \arg \lambda \vert < \frac{\pi}{2} + \frac{\pi}{p-1} - \epsilon\}$, in the case $p=3$ which corresponds to a sextic matrix interaction.}\label{pacman}
\end{figure}

\medskip
One additional remark is that in \cite{KRS,KRS1} we only prove analyticity and Borel-LeRoy summability inside a pacman domain  like Figure \ref{pacman} (see \cite{Wat,Har}). For this article,
we extend to the  more up-to-date cardioid domain of Nevanlinna-Sokal \cite{Nev,Sokal,CaGrMa}.

\medskip\noindent
{\bf Acknowledgement} We would like to thank T. Krajewski, L. Ferdinand, R. Gur\u{a}u, P. Radpay and V. Sazonov  for comments on the present work when we were in some preliminary stage and simply for expressing some interest and motivating us to pursue. We thank Fabien Vignes-Tourneret to have corrected mistakes in Eq. (7) and Eq. (13) of Section 2 of the arXiv 2305.08399. We also acknowledge the support of the CEA-LIST through the  chair ``Artificial Intelligence and Complexity".

\section{The model} 

In this paper, $\cH$ is the Hilbert space $\cH=C^N$,  
$\Tr $ always means \emph{the trace on $\C^{N}$}, $\Tr_\otimes$ always  means \emph{the trace on $\C^{N\times N}$}, and $\bbone_\otimes$ always  means  the $N^2 \times N^2$ matrix whose all eigenvalues are $1$. 

Consider a complex square matrix model with stable interaction of order $2p$, where $p\ge 2$ is an integer 
which is fixed through all this paper.
We assume the reader is reasonably familiar with the notations of \cite{GuKra,KRS,KRS1}
and with Appendix B of the book \cite{guruau2017random}.
Let us recall some basics of our LVR  in the scalar and $d=0$ case \cite{rivasseau2018loop}. 
One of the key elements of the
LVR construction is the Fuss-Catalan numbers of order $p$, which we denote by $C_{n}^{(p)}$, and their generating function $T_p$ \cite{FussCatalan}.
This generating function $T_p$ is defined by
\begin{equation}
T_p(z) = \sum_{n=0}^\infty C_{n}^{(p)} z^n . \label{gencat1}
\end{equation}
It is analytic at the origin and obeys the algebraic equation 
\begin{equation}\label{gencatalan}
zT_p^p(z) -T_p(z) +1 =0 .
\end{equation}

\medskip
In the case $p=3$ the LVR is somewhat simplified; the Fuss-Catalan equation is 
\bee
zT_3^3(z) -T_3(z) +1 =0 , \label{cardanomain}
\ee
which is soluble by radicals. We give in \cite{rivasseau2018loop}, section VI.2,  the details derived from Cardano's solution.

We shall only present our main result for \emph{complex square matrices} in a perturbation $(M M^\dagger )^p $.
In a simplification with respect to \cite{KRS}, we consider only square matrices.
The generalisation to other cases, for instance rectangular complex matrices, or Hermitian matrices, or real symmetric matrices, 
is not too difficult for someone who is familiar of \cite{KRS,KRS1}.
\footnote{For practical applications such as data analysis, the case $p=3$ seems  to be the main one and it is
interesting  to treat the case of \emph{real symmetric} matrices and  \emph{rectangular matrices}.} 

To motivate the introduction of function $T_p$, let us first briefly recall
how the loop vertex representation (LVR) works in the simple scalar case
$N= 1$  \cite{rivasseau2018loop}. In this case, the partition function is simply
\bee
Z(\lambda, 1) = \int dz d\bar z e^{-z\bar z - \lambda (z \bar z)^p}.
\ee
The LVR in this case simply changes variable such that the original action becomes Gaussian; hence,
$ z\bar z + \lambda (z\bar z)^p = w\bar w$. This can be done by choosing
$  \bar w = \bar z, w = z + \lambda (z\bar z)^{p-1} z $. This of course will cost us a Jacobian:
\bee
\frac{\partial (w, \bar w) }{\partial (z , \bar z)}= 1 + p\lambda (z\bar z)^{p-1}.
\ee
Using $z\bar z= w\bar w T_p[-\lambda (w\bar w)^{p-1}]$, we rewrite the partition function as
\bee
Z(\lambda, 1) = \int dw d\bar w e^{-w\bar w-\log \big[1+p\lambda (w\bar w T_p (-\lambda (w\bar w)^{p-1}))^{p-1}\big] }. 
\ee
The derivatives of the log are uniformly bounded in $w, \bar w$ because 
\bee
1- z[T_p(z)]^{p-1} 
\ee has only one cut in the complex plane, which one can avoid by tweaking the phase of $\lambda$.
This allows to control the expansion for $\log Z$.

The case  {\it without sources} of the partition function and its logarithm 
has been treated in \cite{KRS}. Therefore in this paper 
we are dealing with the case  {\it with sources fields $J$}. The source $J$ is itself a $N\times N$ complex matrix and $J^{\dagger}$  is its adjoint. Since $p$ is fixed we omit the subscript $p$ from $T$
when no confusion is possible. 

\begin{definition}
The measure $dM$ and the action $S(M) $ are defined by
\bea
dM&:=&\pi^{-N^2}\prod_{1\leq i,j\leq N}d\text{Re}(M_{ij})d\text{Im}(M_{ij}) \,  ,\label{Z0}
\\
S(M, M^\dagger)  &:=& \Tr \{ M M^\dagger +\lambda (M M^\dagger)^p\} .
\label{Z2}
\eea
The model that we consider in this paper but \emph{without sources} has partition function 
\bea
Z(\lambda, N) &:=& \int\, dM   \, e^{- NS(M, M^\dagger )}.\label{Z1}
\eea
The same matrix model \emph{but with sources} is defined by:
\bea
Z(\lambda,N,J)&=&\frac{\int dMe^{-  NS(M, M^\dagger) +N\Tr(JM^{\dagger})+
N\Tr(MJ^{\dagger})}}{ \int\, dM   \, e^{- NS(M, M^\dagger) }}\label{Z30} 
\eea
where $d M$ and $S(M, M^\dagger) $ are defined in \eqref{Z0}-\eqref{Z2}. 
\end{definition}

For any $N$ by $N$ square matrix $X$ we define the matrix-valued function
\bea
A(\lambda,X)&:=& X T_p(- \lambda X^{p-1})\, ,\label{Adef}
\eea
so that from \eqref{gencatalan}
\begin{equation}
X = A(\lambda,X) + \lambda A^p (\lambda ,X)\,.
\label{AEq}
\end{equation}
We often write $A(X)$ for $A(\lambda,X)$, or even simply $A$, when no confusion is possible.
Next we define an $N$ by $N$ square matrix $X_l$ and an $N$ by $N$ square matrix $X_r$ through
\bea X_l &:=&  M M^\dagger, \quad X_r :=  M^\dagger M . \label{defX} 
\eea
Crucially these two matrices have the same trace, therefore we simply call them $\Tr X$:
\bea \Tr X_l =  \Tr X_r = \Tr X. \label{defX1} 
\eea

For the partition function and its logarithm {\it without sources} the following proposition holds:
\begin{proposition} \label{Pro1} In the sense of \emph{formal power series in} $\lambda$
\bea
Z (\lambda, N) &=& \int\, dM \, \exp\{-N \Tr X + {\cal S}\},
\label{Zgood}
\eea
where ${\cal S}$, the \emph{loop vertex action}, is  
\bea
{\cal S} &=& - \Tr_\otimes \log\Big[\bbone_\otimes+ \lambda \sum_{{\mathfrak k} = 0}^{p-1} A^{\mathfrak k} (X_l) \otimes A^{p-1-{\mathfrak k} }(X_r)\Big]\label{Zgood1}
\\&=& -\Tr_\otimes \log \big[\bbone_\otimes +\Sigma (\lambda,X) \big],
\label{Zgood10}
\eea
where $\Sigma (\lambda,X) := \lambda \sum_{{\mathfrak k}  = 0}^{p-1} A^{\mathfrak k} (X_l) \otimes A^{p-1-{\mathfrak k} }(X_r)$. 
In \eqref{Zgood1} the matrix $A^{\mathfrak k} (X_l)$ acts on the left index of $\cH \otimes \cH$
and the matrix $A^{p-1-{\mathfrak k} }(X_r)$ acts on the right index of $\cH  \otimes \cH$.
\end{proposition}
\prf
This proposition is  proved in \cite{KRS}, Section 2, under the name of Theorem 2.1. 
One proof in \cite{KRS} is by performing a change of variables $M \to P$. $P$ is again an $N$ by $N$ square matrix.
In a manner similar to \cite{KRS} we write $Y := P P^\dagger $, and define $P(M)$ through the implicit function 
formal power series equation $X:= A(Y)$. 
\bea
S(M,M^\dagger) =  \Tr (X +\lambda X^p) =   \Tr [A(Y) +\lambda A^p (Y)] =  \Tr Y ,
\eea
hence it becomes the ordinary Gaussian measure on $P, P^\dagger$.
The new interaction lies therefore entirely in the \emph{Jacobian} of the $M \to P$ transformation.
This transformation can be written more explicitly as
\bea
M &:=& P P^\dagger T_p(-\lambda (P P^\dagger)^{p-1}) (P^\dagger)^{-1} = A(PP^\dagger)(P^\dagger)^{-1},\label{eq100}\\ M^\dagger &:=& P^\dagger\,. 
\eea
Taken into account \cite{KRS} Section 2,  Proposition \ref{Pro1} hold.
\medskip
\qed
\medskip

In terms of $X$ and $M$, equation  \eqref{Z30} (after the change of variables  $P \to M$ in \eqref{eq100}) write 
\bee
Z(\lambda, N, J) := \frac{\int\, d M e^{-N \Tr  (X) +  {\cal S} +
N\Tr(JM^{\dagger})+ N\Tr( A(MM^\dagger)(M^\dagger)^{-1} J^{\dagger})}}
{\int\, d M e^{-N \Tr  (X) +  {\cal S}}},
\label{ZAsq00}
\ee
where ${\cal S}$ does not depend on the sources. 
The quantity $Z(\lambda, N,J)$ may be written in
terms of a Gaussian measure whose covariance is $N^{-1}$:
\bea d\mu (M)  &=&  d M e^{ -N \tr M^\dagger M }, \label{defmu1}\\ 
Z(\lambda, N,J) &=& \frac{\int d\mu (M) e^{ {\cal S}(J,J^\dagger; \lambda,N)+{\cal S}}}{\int d\mu (M)e^{{\cal S}} }. \label{defmu2}
\eea
${\cal S}(J,J^\dagger; \lambda,N)$ depends on the sources and is 
\bea
{\cal S}(J,J^\dagger; \lambda,N) = N\Tr(JM^{\dagger})+ N\Tr( M M^\dagger T_p(-
\lambda MM^\dagger)(M^\dagger)^{-1}J^{\dagger}) ,
\label{eqZ2}
\eea
where $T_p(x,u)$ is the generating function of rooted $p$-ary trees with a factor $x$ per internal vertex and $u$ per leaf.

Next we shall define a mathematical expression for the cumulants.     

\begin{definition}
\label{cumulantsdef}
The cumulant of order $2{\mathcal K}$ is:
\bea
\hskip-.6cm\label{cum01}
\mathfrak{K}^{{\mathcal K}}(\lambda,N )\!&:=&\!\Big[
\frac{\partial^{2}}{J^{\ast}_{a_{1}b_{1}} J_{c_{1}d_{1}}}\cdots
\frac{\partial^{2}}{J^{\ast}_{a_{{\mathcal K}}b_{{\mathcal K}}} J_{c_{{\mathcal K}}d_{{\mathcal K}}}}
\log {\cal Z}(\lambda, N,J)\Big]_{J=0}, 
\eea
where
\bea \label{diffnor}
\log {\cal Z} (\lambda, N,J)= \frac{1}{N^2} \log Z(\lambda, N,J) 
\eea
and $J_{ab}^{\ast}$ is the complex conjugate of $J_{ab}$, so that $(J^{\dagger})_{ab}=J^{\ast}_{ba}$.
\end{definition}

\medskip
Note that all the derivatives of $\log {\cal Z}$ which are not of this form vanish. 
In the case $p=2$, which is treated by \cite{GuKra}, the order 2 cumulant is proportional to 
$\langle M_{ab}M^{\ast}_{cd}\rangle-\langle M_{ab}\rangle\langle M^{\ast}_{cd}\rangle$
whereas, if we transpose,  it is {\it not} the same thing: $\langle  M_{ba}M^{\ast}_{cd}\rangle  -  \langle M_{ba}\rangle 
\langle M^{\ast}_{cd}\rangle$ is subtly different from $\langle M_{ab}M^{\ast}_{cd}\rangle-\langle M_{ab}\rangle\langle M^{\ast}_{cd}\rangle$.

Any quantity $F$ in quantum field theory 
which is an integral over a Gaussian measure $\int_C d \mu (\phi) \ F(\phi )$ 
can be combinatorially represented as a sum over the set $\gF$
of \emph{oriented forests}\footnote{Oriented forests simply distinguish edges $(i,j)$
and $(j,i)$ so have edges with arrows. It allows to distinguish below between operators 
$\frac{\partial}{\partial M^\dagger_i}\frac{\partial}{\partial M_j^{\textcolor{white}\dagger}}$ and $\frac{\partial}{\partial M^\dagger_j}\frac{\partial}{\partial M_i^{\textcolor{white}\dagger}}$.}
by applying the BKAR formula \cite{BK,AR1}.
For readers who want to look further into BKAR formula and oriented forests, ordered or not, see \cite{RiZh1,RiTa,GuKra}.

In this context, where the Gaussian measure is $d \mu (M) $ and the
covariance is $C= \frac1N$, we start by replacing the covariance by $C_{ij}(x) =\frac{x_{ij} + x_{ji}}{2} \frac1N $ evaluated at $x_{ij} = 1$
for $i \neq j$ and $C_{ii}(x) = \frac1N \; \forall i$. Then the Taylor BKAR formula for oriented forests  $\gF_n$ on $n$ labeled vertices  yields
\bea 
&&\hskip-3.7cm F(M) =\sum_{n=0}^\infty\frac{1}{n!}\;\;\sum_{\cF\in \gF_n}\; \int dw_\cF\  \partial_\cF \; 
\int\, d\mu_{C\{x_{ij}^\cF\}} (M)\,F_n(M)\;  \Big|_{x_{ij} = x_{ij}^\cF (w)},
\label{LVE1a}\\
\text{where }\int dw_\cF &:=& \prod_{(i, j) \in \cF} \int_0^1dw_{ij} \; ,  \quad
\partial_\cF  := \prod_{(i, j) \in \cF} \frac{\partial}{\partial x_{ij}} \; ,\label{LVE2a}\\
x_{ij}^\cF(w) &:=& \left\{ 
\begin{array}{c}
\hspace{-1.4cm}\text{inf}_{(k,l)\in P^\cF_{i\leftrightarrow j}} w_{kl}~~~~~~~~~~~ \text{if}~ P^\cF_{i\leftrightarrow j}~ \text{exists}\,, \\ 
0~~~~~~~~~~~~~~~~~~~~~~~~~~~~\text{if}~ P^\cF_{i\leftrightarrow j}~ 
\text{does not exist}\, .\end{array}\right.\label{LVE4a}
\eea
In this formula $w_{ij}$ is the weakening parameter of the edge $(i,j)$ of the forest, 
and $P^\cF_{i\leftrightarrow j}$ is the unique path in $\cF$ joining $i$ and $j$ when it exists.

Remember that a main property of the forest formula is that the symmetric $n$ by $n$ matrix 
$C\{x^\cF_{ij} \}=\frac{x^\cF_{ij} (w)+ x^\cF_{ji}(w)}{2} \frac1N $
is positive for any value of $w_{kl}$, hence the Gaussian measure $d\mu_{C\{x^\cF_{ij}\}} (M) $ is well-defined. 
Since the fields, the measure and the integrand are now 
factorized over the connected components of $\cF$, its \emph{logarithm}
is easily computed as exactly the same sum but restricted to the spanning trees. 

\section{LVR Amplitudes of Combinatorial Maps}
\footnote{This section is essentially a condensed version of the corresponding section of \cite{GuKra}.}
For the task of computing 
$\mathfrak{K}^{k}(\lambda,N) $, we have to introduce combinatorial maps, a refinement of the usual Feynman graphs. A combinatorial map is a graph with a distinguished cyclic ordering of the half edges incident at each vertex. Combinatorial maps are conveniently represented as \emph{ribbon graphs} whose vertices are disks and whose edges are ribbons (allowing one to encode graphically the ordering of the half edges incident at a vertex). When applied to cumulants, it is based on combinatorial maps with cilia. 
A \emph{cilium} is a half edge hooked to a vertex.
We denote $k(G)$, $v(G)$, $e(G)$, $f(G)$ and $c(G)$ the cilia,
vertices, edges, faces and \emph{corners} of $G$ \footnote{Attention: in a simplification to the notations of \cite{GuKra},
we shall no longer use in this paper the distinction between $k(G),v(G),e(G),f(G),c(G)$ and their \emph{numbers} $|k(G)|,|v(G)|,|e(G)|,|f(G)|,|c(G)|$.}
A corner of $G$ is a pair of consecutive half edges attached to the same vertex. 
The faces of $G$ are partitioned between the faces which do not contain any cilium (which we sometimes call internal faces) and the ones which contain at least a cilium, 
which we call \emph{broken faces}.  We denote $b(G)$ the set of broken faces of $G$. 
Each broken face corresponds to a puncture in the Riemann surface in which $G$ is embedded.

\begin{figure}[!htb]
\centering\includegraphics[width=0.5\linewidth]{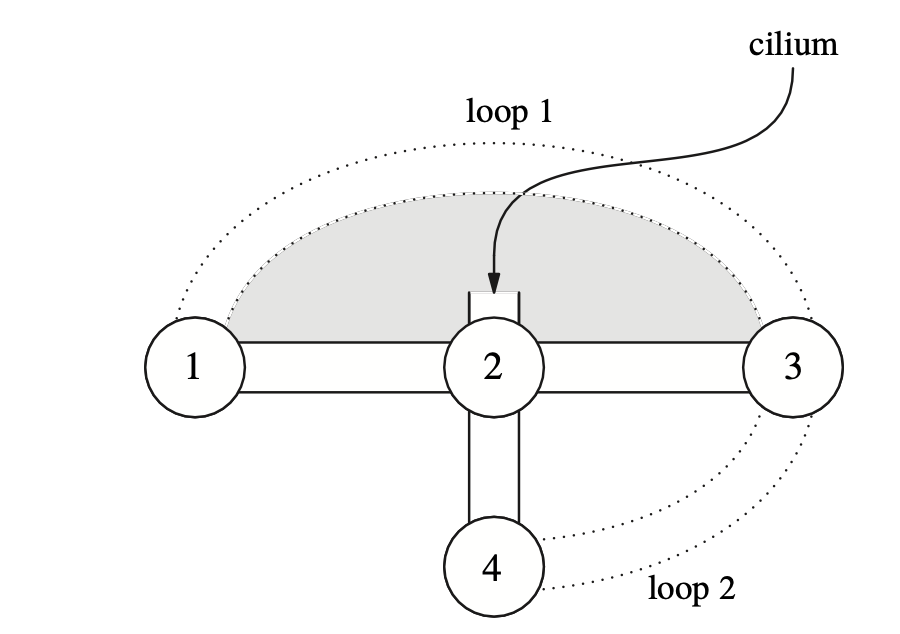} 
\caption{A LVR graph in the case $p=2$ with one cilium and one broken face 
(coloured in grey, courtesy of \cite{GuKra}).}\label{Fig3}
\end{figure}

The Euler characteristic of the graph $G$ is:
\begin{equation}
\chi(G)=v(G)-e(G)+f(G)-b(G)=2-2g(G)-b(G),
\label{Euler:eq}
\end{equation}
 where  $g(G)$ is the genus of the graph $G$. 

\begin{definition}[LVR graphs and trees] 
A LVR graph $(G,T)$ is a connected ribbon graph $G$ with labels on its vertices having furthermore:
\begin{itemize}
\item a distinguished spanning tree $T\subset G$,
\item a labeling of the edges of $G$ not in $T$,
\item at most one cilium per vertex. 
\end{itemize}
A LVR tree is a graph such that the set $l(G,T):=e(G)-e(T)$ is empty, so $(G,T)=(T,T)$.
\end{definition}

We associate to every LVR graph $(G,T)$ \emph{its amplitude} ${\cal A}_{(G,T)}(\lambda,N,J)$.
We emphasize here that the following definition of amplitudes is almost the same that in \cite{GuKra},
but is subtly \emph{different for the LVE and for the LVR}; we have  to replace the intermediate field
(denoted by $A$ in \cite{GuKra})
by a {\it family of functions $X_l, X_r$} depending only on the fields $M_i,M_i^{\dagger}$ by eq. \eqref{defX}
but {\it having the same trace}, see \eqref{defX1}. 
This point is subtle and we want to explain in more detail. First we want to make the following definition : 
\begin{definition}
\bea
\label{Sigmadef1}
\Sigma (\lambda,M ) &:=&  \lambda \sum_{{\mathfrak k} = 0}^{p-1} A^{\mathfrak k}(MM^\dagger) \otimes A^{p-1-{\mathfrak k}}(M^\dagger M).
\eea
where $A$ is defined by \eqref{Adef}.
\end{definition}

\begin{definition}
We define the \emph{amplitude} of LVR graph $(G,T)$, ${\cal A}_{(G,T)}(\lambda,N)$, by 

\bea
\label{LVRamplitude}
{\cal A}_{(G,T)}(\lambda,N,J)&=&\frac{(-\lambda)^{e(G)}N^{v(G)-e(G)}}{v(G)!}
\mathop{\int}\limits_{1\geq s_{1}\geq\cdots\geq s_{|L(G,T)|}\geq 0}\,\prod_{e\in L(G,T)}ds_{e}
\nonumber\\&&
\mathop{\int}\limits_{[0,1]} \prod_{e\in E(T)} dt_{e}
\left( \prod_{e=(i,j)\in L(G,T)}\mathop{\text{inf}}\limits_{e'\in P_{i\leftrightarrow j}^{T}} t_{e'}  \right) 
\int d\mu_{s_{|L(G,T)|}C_{T}}
\nonumber\\&&
\prod_{f\in f(G)}\Tr\bigg\{\mathop{\prod}\limits_{c\in\partial f}^{\longrightarrow}
\big[\bbone_\otimes   +\Sigma (\lambda,M_{i_{c}})\big]^{-1} (JJ^{\dagger})^{\eta_{c}}\bigg\} \; ,
\eea
where:
\begin{itemize}
\item ${\displaystyle\mathop{\prod}^{\longrightarrow}}$ is the oriented product around the corners $c$ 
on the boundary $\partial f$ of the face $f$,
\item $i_c$ is the label of the vertex the corner $c$ belongs to. 
\item $\eta_{c}=1,0$ depending on whether $c$ is followed by a cilium or no cilia, 
\item the Gaussian measure $\int d\mu_{C\{x_{ij}^\cT\}} (M)$ can also be written as the differential operator:
\bea
&& \hskip-1.8cm\int d\mu_{C\{x_{ij}^\cT\}}(M) F \{ M_i, M_i^{\dagger }\}
= \left[  e^{\frac{x_{ij}^T + x_{ji}^T}{2}\frac{\partial}{\partial M_i} \frac{\partial}{\partial M^{\dagger}_j}} \; 
F\Big\{ M_i, M^{\dagger}_i \Big\} \right]_{\{M_i\}=0 },
\eea
\item and $\Sigma (\lambda,M)$ is defined by \eqref{Sigmadef1}.
\end{itemize}
\end{definition}
The propagator is the same for $M_{i_{c}}$ and  $M^\dagger_{i'_{c}}$. The sources $J$ and $J^\dagger$ are also the same, but the tree, the loops and the Feynman graphs are different in the LVR than in the LVE. For $p\ge 3$
there is {\it no intermediate field representation as in \cite{GuKra}}, 
there is only the direct representation with $M$ and $M^\dagger$, with one $M$ and one $M^\dagger$ that plays a special role. Fortunately there is also {\it a  cusp or half-edge $\sqcup$} to distinguish the sources  $J$ and $J^\dagger$ and the $M_{i_{c}}$ and $M^\dagger_{i'_{c}}$.
 
\begin{figure}[!t]
\begin{center}
{\includegraphics[width=12cm]{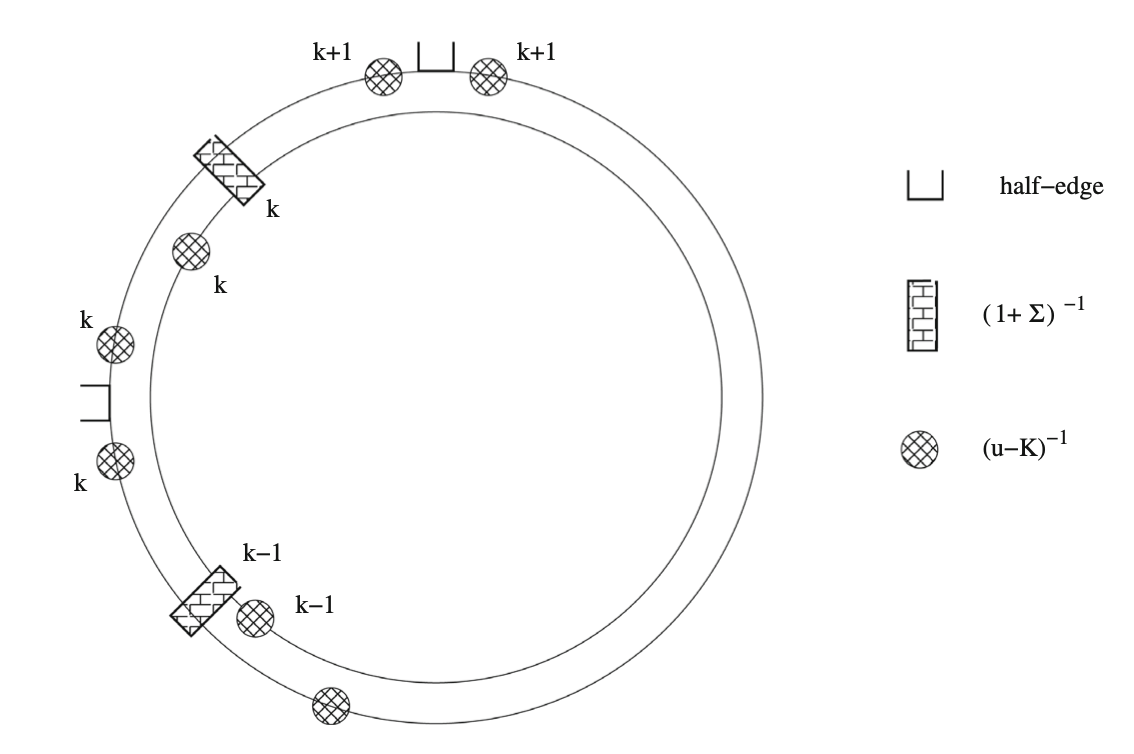}}
\end{center}
\caption{A vertex with some of its corner operators. The label $k$ indicates the corresponding contour variable. 
See the upper left corner; between the two half-edges $\sqcup$ symbols it contains three $(u-K)^{-1}$ operators with indices $k$, $k$ and $k+1$ and (crucially because it connects the two circles) {\it one resolvent $(1+\Sigma(\lambda,M))^{-1}$ with index $k$}. 
For the definition of $u$ and $K$ see \cite{KRS,KRS1}.
The first $\frac{\partial}{\partial K}$ derivative 
is a bit special as it destroys forever the logarithm in $\LV$ and gives
$\bigg[\frac{\partial}{\partial K}\bigg]  \Tr_\otimes \log \big[\bbone_\otimes +\Sigma  \big] = 
\big[\bbone_\otimes +\Sigma \big]^{-1} \frac{\partial \Sigma }{\partial K}$. 
To compute $\frac{\partial \Sigma }{\partial K}$
we can use holomorphic functional matrix calculus as in \cite{KRS}
to  write $\big[\bbone_\otimes  +\Sigma_g(K) \big]^{-1} 
= \frac{ K\otimes \mathbf{1} - \mathbf{1} \otimes K}
{H \otimes \mathbf{1} - \mathbf{1} \otimes H}$,
with $\Sigma_g (K)  =  \oint_{\Gamma} du\; [h_g(u)-u]  \frac{1}{u-K}  \otimes  \frac{1}{u-K}$, where $g$ and $h$ 
are defined in \cite{KRS,KRS1}
and where the contour $\Gamma$ is any  contour enclosing 
the spectrum of $K$. The final tree amplitude will be obtained later by gluing these $\sqcup$ symbols 
for $M$ and $M^\dagger$ along the edges of the trees.}
\label{corneropfig}
\end{figure}

 There is also a vertex with its corners, see Figure \ref{corneropfig}. 
To perform a computation of a vertex, we use  the  $\sqcup$ symbols as in \cite{KRS}
for the pairing of the regular $M_{i_{c}}$ and $M^\dagger_{i'_{c}}$, and for the sources $J$ and $J^\dagger$, there in no pairing: one simply glue the half-edge $\sqcup$ with the corresponding $J$ or $J^\dagger$.

\medskip
If the graph $(G,T)$ has $k$ cilia we use for its amplitude the notation ${\cal A}^k_{G,T}(\lambda,N,J) $;
if the graph $(G,T)$ is reduced to a tree we use the shorthand notation ${\cal A}^k_{T}(\lambda,N,J) $ instead of 
${\cal A}^k_{(T,T)}(\lambda,N,J)$. The amplitude simplifies drastically in this case:
one trace is obtained (as trees have only one face). Hence
\bea
\label{LVEamplitudefortree}
{\cal A}^k_{T}(\lambda,N,J)&=&\frac{(-\lambda)^{e(T)}N^{
v(T)-e(T)
}}{v(T)!}\\ \nonumber &&
\int d\mu_{C\{x_{ij}^\cT\}} (M)\Tr\bigg\{\mathop{\prod}\limits_{c\in\partial f}^{\longrightarrow}
\big[\bbone_\otimes  +\Sigma (\lambda,M) \big]^{-1}(JJ^{\dagger})^{\eta_{c}} \bigg\} \; .
\eea

\begin{figure}[htb]
\begin{center}
\includegraphics[width=6cm]{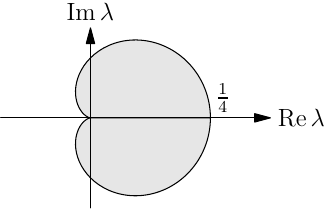}
\caption{Cardioid domain ${\cal C}$ in the complex $\lambda$ plane in the case $p=3$.}
\label{cardioidpic}
\end{center}
\end{figure}

Let ${\cal C}$  be the cardioid domain  
 \begin{equation}{\cal C}=\bigg\{\lambda\in{ \mathbb{C} }
 \quad\text{with}\quad 
| \lambda|<\frac{1}{2(p-1)}\cos^{p-1}\Big({\frac{\arg\lambda}{p-1}}\Big)\bigg\} \; ,
 \label{cardioid00} 
\end{equation}
where we choose the determination $-\frac{\pi}{2}<\frac{\arg\lambda}{p-1}<\frac{\pi}{2}$ of the argument.
This domain is also defined in the Appendix of this article  by ${\cal D}_R$, 
simply change $q\to p-1$, $z\to \lambda$, $R \to \frac{1}{2(p-1)}$.
For the case $p=3$, ${\cal C}$  is pictured in Figure \ref{cardioidpic}.

\section{Results}
\label{section4}
We state an analyticity result and a Borel summability for the constructive expansion of the cumulant of order $2{\mathcal K}$ or {\it J-cumulant}. We recall that all the results of this section  {\bf are valid only for $1\le {\mathcal K}\le {\mathcal K}_{\max}$, where $ {\mathcal K}_{\max}$ is fixed}. Then is our main Theorem.
\begin{theorem}[Constructive expansion for the J-cumulant] \label{th1}
Let $1\le {\mathcal K}\le {\mathcal K}_{\max}$, where $ {\mathcal K}_{\max}$ is fixed. There exists 
$\epsilon_{\lambda}>0$ depending on $\lambda$ such that 
$\mathfrak{K}^{{\mathcal K}}(\lambda,N) $ is given by the following absolutely convergent expansion 
\bea
\mathfrak{K}^{{\mathcal K}}(\lambda,N) &=&
{\cal P}^{\mathcal K}_{n}(\lambda,N,J)+{\cal Q}^{\mathcal K}_{n}(\lambda,N,J) + {\cal R}^{\mathcal K}_{n}(\lambda,N,J)\label{treeexpansion00},\\
{\cal P}^{\mathcal K}_{n}(\lambda,N,J)&=&\sum_{\substack{\substack{G\text{ labeled ribbon graph }\\ \text{ with ${\mathcal K}$ cilia,}} 
\\ e(G)\leq n}} \hskip-.3cm
\frac{(-\lambda)^{e(G)}N^{\chi(G)}}{v(G)!}\hskip-.1cm \prod_{f\in b(G)} \hskip-.1cm\Tr[(JJ^\dagger)^{c(f)}]\label{treeexpansion01} ,\\
\hskip-.3cm {\cal Q}^{\mathcal K}_{n}(\lambda,N,J)&=&\sum_{\substack{\substack{(G,T)\text{ LVR graph }\\ \text{  with ${\mathcal K}$ cilia}}\\ e(T)= n+1}}{\cal A}^{\mathcal K}_{(G,T)}(\lambda,N,J) ,
 \label{treeexpansion02}
 \\{\cal R}^{\mathcal K}_{n}(\lambda,N,J)&=&\sum_{\substack{\substack{T\text{ LVR tree  }\\ \text{ with ${\mathcal K}$ cilia}}\\ e(T)\geq n+2}}{\cal A}^{\mathcal K}_{T}(\lambda,N,J) .
 \label{treeexpansion03}
\eea
This expansion is analytic for any $\lambda\in {\cal C}$ 
and the remainder at order $n$ obeys, for $\sigma$ constant large enough, the analog of \eqref{BorLeRSum0}  
\bee
| {\cal R}^{\mathcal K}_{n}(\lambda,N,J)| =  \big| \mathfrak{K}^{{\mathcal K}}(\lambda,N,J)-\sum_{m=0}^{n}a_{m}(N,J)\lambda^{m} \big|\le \sigma^n\,
[(p-1)n] ! \, | \lambda |^{n+1} ,\label{BorLeRSum01}
\ee
uniformly in $N\in{\mathbb N}^*$, J such that $\|J^{\dagger}J\|<\epsilon_{\lambda}$.
Therefore  it obeys the theorem stated in the Appendix  of this article (Borel-LeRoy-Nevanlinna-Sokal) with $q \to p-1$, $z \to \lambda$, 
$\omega\to \big\{ N,J \big\}$, whenever $N\in{\mathbb N}^*, \|J^{\dagger}J\|<\epsilon_{\lambda}$.
\end{theorem}

\section{Proof of Theorem \ref{th1}.}
\subsection{Strategy}

Because $\mathfrak{K}^{{\mathcal K}}(\lambda,N)$ is a sum of three pieces, Theorem \ref{th1} contains three pieces, respectively indexed by ${\cal P}^{\mathcal K}_{n}(\lambda,N,J)$, 
${\cal Q}^{\mathcal K}_{n}(\lambda,N,J)$ and ${\cal R}^{\mathcal K}_{n}(\lambda,N,J)$,
 So the proof can be decomposed into three parts:
\begin{itemize}
\item the one who concerns ${\cal P}^{\mathcal K}_{n}(\lambda,N,J)$,
\item the one who concerns ${\cal Q}^{\mathcal K}_{n}(\lambda,N,J)$,
\item and the part that concerns  the remainder at order $n$, ${\cal R}^{\mathcal K}_{n}(\lambda,N,J)$.
\end{itemize}
For the first part, the one who concerns
\bee
\sum_{\substack{\substack{G\text{ labeled ribbon graph }\\ \text{ with $2{\mathcal K}$ cilia }}\\ |e(G)|\leq n}} \hskip-.3cm
\frac{(-\lambda)^{|e(G)|}N^{\chi(G)}}{|v(G)|!}\hskip-.1cm \prod_{f\in b(G)} \hskip-.1cm\Tr (JJ^\dagger)^{c(f)}
\ee
 the proof is rather trivial: it is absolutely convergent expansion and this expansion is analytic for any $\lambda\in {\cal C}$ 
since it is is not only analytic but polynomial. 
It remain to prove the two other parts. Here we go.

\subsection{Proof of the part concerning Q}

This part concerns a sum over labeled ribbon graphs with ${\mathcal K}$ cilia of
 LVR amplitudes ${\cal A}_{(G,T)}(\lambda,N,J)$. But in the definition of ${\cal A}_{(G,T)}(\lambda,N,J)$
 given by \eqref{LVRamplitude}, the only part depending on 
 $J$, is $(J^{\dagger}J)^{\eta_{c}}$.
 Since for $1\le {\mathcal K} \leq {\mathcal K}_{max}$,  $(J^{\dagger}J)^{\eta_{c}}$ is evidently bounded by $\|J^{\dagger}J\|^{\mathcal K}$,
 we do not worry about this part. It holds. 

We turn now to the part no longer dependent on $J$,
hence dependent of holomorphic matrix calculus and to contour integrals like \cite{KRS}; therefore
no wonder we are going to make a heavy use of the notations and the results of \cite{KRS}.
It is crucial to make the distinction between 
$\Tr $ and $\Tr_\otimes$. To simplify the notations of this subsection we often forget the dependency on $\lambda$ 
when no confusion is possible. For example we often write simply 
 $\Sigma(M)$ or even $\Sigma$ for $\Sigma(\lambda,M)$
 where $\Sigma(\lambda,M)$ is defined by \eqref{Sigmadef1}. 
 
\begin{figure}[!ht]
\begin{center}
{\includegraphics[width=12cm]{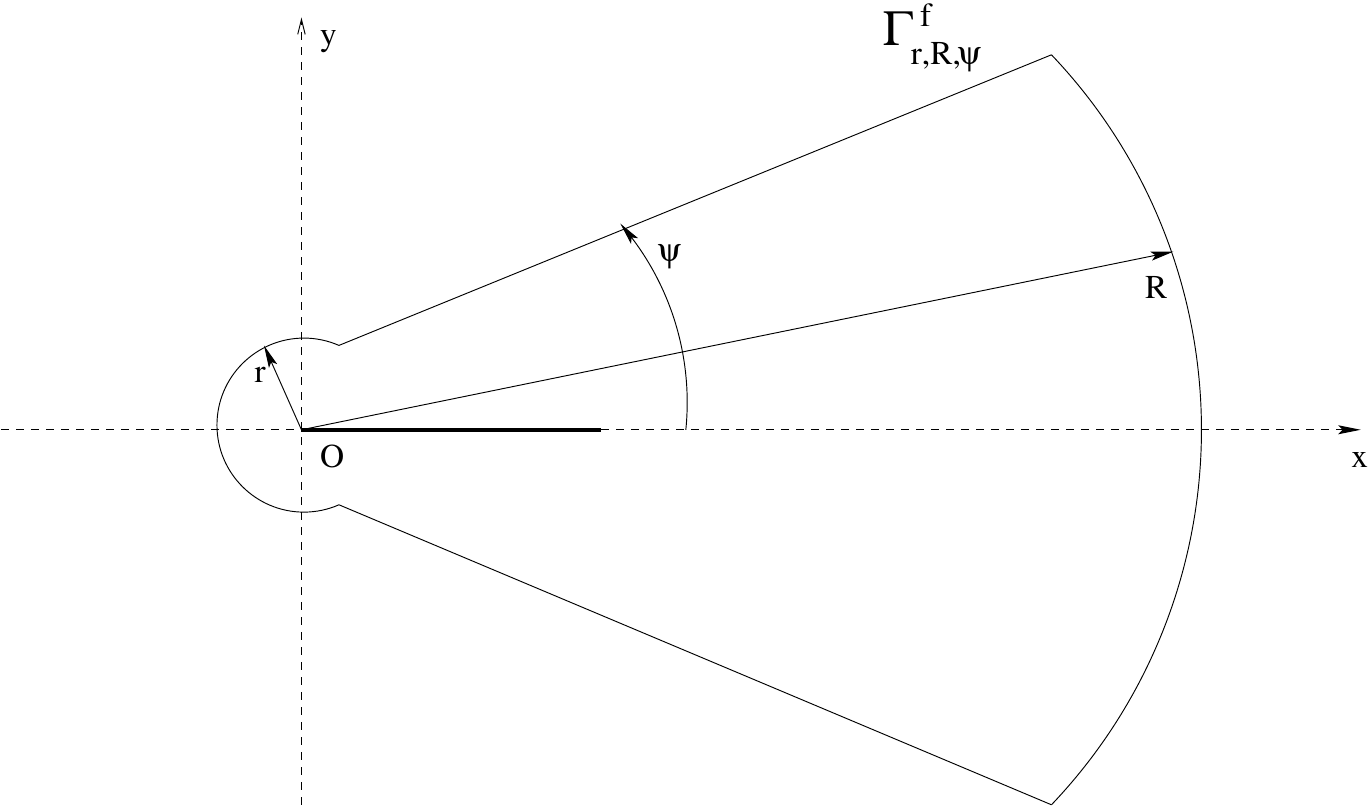}}
\end{center}
\caption{A \emph{finite} keyhole contour $\Gamma^f_{r, R, \psi}$ encircling a segment 
on the real positive $Ox$ axis, which includes the spectrum of $X$.
 The spectrum of $X$ lies on a real axis positive segment, like the one shown in boldface.}
\label{keyholeencirc0}
\end{figure}
 
 Given a holomorphic function $f$ on a domain containing the spectrum of a square  
matrix $X$, Cauchy's integral formula yields a convenient expression for $f(X)$:
\bee
f(X)=\oint_{\Gamma} du\frac{f(u)}{u-X},  \label{cauch0}
\ee
provided the contour $\Gamma$
is a \emph{finite} keyhole contour enclosing all the spectrum of $X$ (see Figure \ref{keyholeencirc0}).

In \cite{KRS} it is established that 
\bee  A(\lambda,X) =  \oint_{\Gamma} du\; a(\lambda,u)   \, \frac{1}{u-X}, \label{cauch1}
\ee
where $a(\lambda, u) = u T_p(-\lambda u^{p-1})$ (see \eqref{gencatalan},\eqref{Adef}).
On the other hand by a useful lemma also proven in \cite{KRS}, we know that
\bee  \frac{\partial A}{\partial X}= 
\big[\bbone_\otimes +\lambda\Sigma (\lambda,X)\big]^{-1} =\big[\bbone_\otimes +\lambda\Sigma \big]^{-1} . \label{resoder}
\ee

Then we can write the matrix derivative acting on a resolvent.
We obtain by wrtiting again the superscript $i$
\bee
\frac{\partial A}{\partial X^{i}} =  \big[\bbone_\otimes +\lambda\Sigma^{i} \big]^{-1} =
\oint_{\Gamma} du\; a(\lambda, u) \frac{1}{u- X^{i}}\otimes  \frac{1}{u- X^{i}}.
 \label{cauchA}
\ee

\begin{figure}[!ht]
\begin{center}
{\includegraphics[width=10cm]{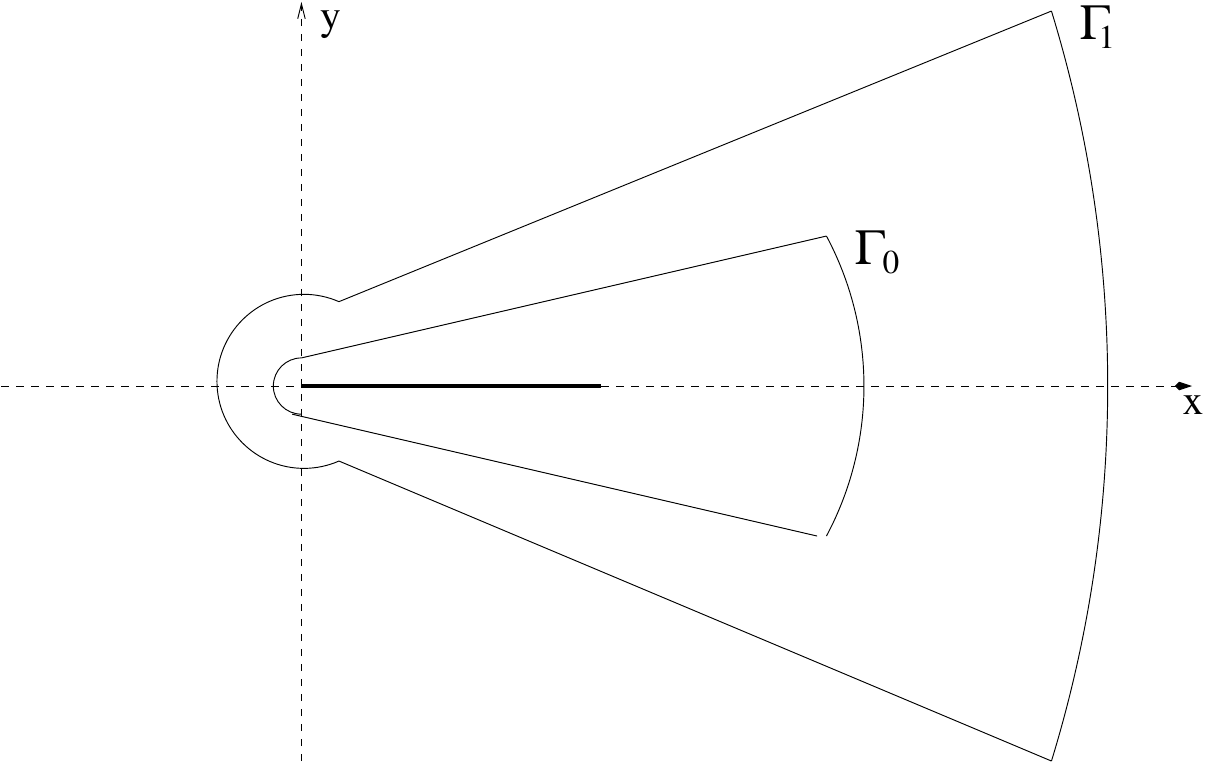}}
\end{center}
\caption{A keyhole contour $\Gamma_1$ encircling a keyhole contour $\Gamma_0$.}
\label{keyholeencirc1}
\end{figure}

Now we reapply the holomorphic calculus, but in different ways\footnote{Our choices below are made in order to 
allow for the bounds of Section \ref{bound}.} depending on the term chosen in the sum over $k$.
\begin{itemize}
\item For $k=0$, we apply the holomorphic calculus to the right  
$ \frac{A^{p-1}(\lambda,X)}{u-X} $
factor, with a contour $\Gamma_2$ surrounding $\Gamma_0$
for a new variable called $v_2$, and we rename $u$ and $\Gamma_0$ as $v_1$ and $\Gamma_1$ (see Figure \ref{keyholeencirc1}),

\item for $k=p-1$, we apply the holomorphic calculus to the left $ \frac{A^{p-1}(\lambda,X)}{u-X} $ factor, with a contour $\Gamma_2$ surrounding $\Gamma_0$
for a new variable called $v_2$, and we rename $u$ and $\Gamma_0$ as $v_1$ and $\Gamma_1$; we obtain a contribution identical to the previous case,

\item  in all other cases, hence for $1\le k \le p-2$, we apply the holomorphic calculus both to left and right factors in the tensor product,
with two variables $v_1$ and $v_2$ and two equal contours $\Gamma_1$ and $\Gamma_2$ enclosing enclose the contour $\Gamma_0$.

\end{itemize}

Recall that the first $\frac{\partial}{\partial X}$ derivative 
is a bit special as it destroys forever the logarithm in ${\cal S} (\lambda,X)$ and gives
\bee \bigg[\frac{\partial}{\partial X}\bigg]  \Tr_\otimes \log \big[\bbone_\otimes +\Sigma  \big] = 
\big[\bbone_\otimes +\Sigma \big]^{-1} 
\frac{\partial \Sigma }{\partial X} \;. \label{firstder}
\ee

Recall $X_l = MM^\dagger$, $X_r= M^\dagger M$, \eqref{defX1} and recall, in \eqref{LVRamplitude}, the part no longer dependent on $J$:
\bee
\frac{(-\lambda)^{e(G)}N^{v(G)-e(G)}}{v(G)!}
\int dw_{T} \partial_{T} \int d\mu_{C\{x_{ij}^\cT\}}(M) 
\prod_{f\in f(G)}\Tr \mathop{\prod}\limits_{c\in\partial f}^{\longrightarrow}
\big[\bbone_\otimes   +\Sigma (\lambda,M)\big]^{-1} .\label{LVRam01}
\ee
Let us for the moment concentrate about the part depending on $\Tr \mathop{\prod}\limits_{c\in\partial f}^{\longrightarrow}
\big[\bbone_\otimes   +\Sigma (\lambda,M)\big]^{-1}$
Defining the loop resolvent
\bee \cR (v_1, v_2, M, M^\dagger) :=   \label{resolv1}
 \Big[\Tr   \frac{1}{v_1-MM^\dagger} \Big]  \Big[\Tr   \frac{1}{v_2-M^\dagger M} \Big] ,
\ee
we obtain
\bea  \frac{\partial {\cal S}}{\partial \lambda} &=&- \oint_{\Gamma_1} dv_1 \oint_{\Gamma_2} dv_2  \Big\{ \oint_{\Gamma_0} du\; a(\lambda, u) \sum_{{\mathfrak k} = 1}^{p-2} 
\frac{\partial_\lambda [\lambda a^{\mathfrak k} (\lambda,v_1) a^{p-{\mathfrak k} -1}(\lambda,v_2)]}{(v_1 - u)(v_2 - u)} \, \nonumber \\
&& \qquad +\; 2 a(\lambda, v_1) \frac{ \partial_\lambda \big[ \lambda a^{p-1}(\lambda,v_2) \big]  }{v_1 - v_2}  \Big\}\cR (v_1, v_2, M, M^\dagger).
 \label{niceder1}
\eea
In the manner of \cite{KRS} we can now
commute the functional integral and the contour integration. This results in
\bea
&&
\int dw_{T} \partial_{T} \int d\mu_{C\{x_{ij}^\cT\}}(M) 
\prod_{f\in f(G)}\Tr \mathop{\prod}\limits_{c\in\partial f}^{\longrightarrow}
\big[\bbone_\otimes   +\Sigma (\lambda,M)\big]^{-1} \\&&
=\sum_{T} \int \{dw_{T} dt du dv\} \Phi_n \nonumber
\int d\mu_{C\{x^T\}} (M) \partial_T^M   \cR^{k}_n \Big|_{x_{ij} = x_{ij}^T (w)} ,
\label{LVE4000}
\eea
where 
\bea
\partial_T^M &:=& \prod_{(i, j) \in T} \Tr_{\otimes}\Big[
 \frac{x_{ij}^T + x_{ji}^T}{2}\frac{\partial}{\partial M_i} \frac{\partial}{\partial M^{\dagger}_j} \Big],\label{LVE4001}\\
\cR^{k}_n &:=& \prod_{i = k+1}^n  \cR ( v^i_1 , v^i_2, M_i, M_i^\dagger) , \label{LVE4003}
\eea
and the symbol $\int \{dw_T  dt du dv\} \Phi_n $ stands for
\bea \int  \{dw_T dt du dv \} \Phi_n     &=& \prod_{i ,j\in T}\int_0^1 dw_{ij}
\prod_{i = 1}^n \Bigg[  \int_0^\lambda dt^i \oint_{\Gamma^i_1} dv^i_{1} \oint_{\Gamma^i_2}  dv^i_{2} 
 \nonumber\\&&\Big\{\oint_{\Gamma^i_0}  du^{i}  \phi (t^i, u^i , v^i_1 , v^i_2) +\psi (t^i, v^i_1 , v^i_2)  \Big\}\Bigg].
 \label{LVE4004}  
\eea

The trace $\Tr_{\otimes}$ in \eqref{LVE4001} can also be thought 
as two  independent  traces $\Tr$ associated to ordinary loops (hence the name ``loop vertex representation"). 
In \eqref{LVE4000} they are only coupled through the scalar factors of \eqref{LVE4004}.
The nice property of this LVR representation is that it does not break the symmetry between the two factors of the tensor product in \eqref{cauchA}. 
 
The condition on the contours
 $\Gamma^f_{r_j, \psi_j, R_j}$ for $j= 0,1, 2$, can be written 
 \bea 
 &&0< \psi_0 < \min (\psi_1, \psi_2) \le   \max (\psi_1, \psi_2) < \delta ,
 \\&&
 0< r_0 < \min (r_1, r_2); \quad \Vert MM^\dagger \Vert  + 1  \le R_0 < \min (R_1, R_2).
 \eea  
${\cal S}$ is \emph{not uniformly bounded in $MM^\dagger$} but grows logarithmically at large $\Vert MM^\dagger \Vert $.
However it \emph{fully disappear in the LVR formulas below}, because these formulas  do not use ${\cal S}$ but derivatives of ${\cal S}$
with respect to the field $M$ or $M^\dagger$. Hence we may use
\emph{infinite} contours  $\Gamma^\infty_{r, \psi}$ which are completely independent of $\Vert MM^\dagger \Vert $ \cite{KRS}.

The outcome of applying  $\partial^M_T$ to $\cR^{k}_n $
is a bit difficult to write, but the combinatorics has been treated in \cite{KRS}. For a single loop 
the Fa\`a di Bruno formula allows to write this outcome as a sum over a set  $\Pi^{q, \bar q}_r$ of Fa\`a di Bruno  terms, 
each one of these with a factor 1:
\bea
\frac{\partial^r }{\partial M_1 \cdots \partial M_q  \partial M^\dagger_1  \cdots \partial M^\dagger_{\bar q} }
\frac{1}{v-X} 
\!&=&\!\hskip-.1cm \sum_{\pi \in \Pi^{q, \bar q}_r} \;\Tr\Big[O^\pi_0 \sqcup O^\pi_1 \sqcup \cdots \sqcup O^\pi_r \Big]. \label{faasum}
\eea
In the sum \eqref{faasum} there are exactly $r$  symbols $\sqcup$, separating $r+1$ corner operators $O^\pi_c$.

The result of this computation is obtained by identifying the two ends of each pair of $\sqcup$ symbols along each edge of $T$. This pairing of the $2n-2$ $\sqcup$ symbols then exactly glue the $2n$ traces of the tensor products present in the $n$ vertices into $n+1$ traces.
These corner operators can be of four different types, either resolvents $\frac{1}{v-X}$, $M$-resolvents $\frac{1}{v-X} M $, $M^\dagger$-resolvents 
$M^\dagger \frac{1}{v-X}  $,
or the identity operator $\bbone$. We call $r_\pi$, $r^M_\pi$, $r^{M^\dagger}_\pi$ and $i_\pi$ the number of corresponding operators in $\pi$.
By a lemma proven in  \cite{KRS} we know
\bee \vert \Pi^{q, \bar q}_r \vert \le   2^r  r!, \quad
r_\pi   =   1  +  i_\pi    ,  \quad   r^M_\pi  +r^{M^\dagger}_\pi = r - 2 i_\pi .
\label{faacomb}
\ee 

Applying \eqref{faasum} at each of the two loops of each loop vertex, we get for any tree $T$
\bee  \partial_T^M  \Tr  \frac{1}{v-X} 
= \prod_{i=1}^n \Big\{ \prod_{j= 1}^2 \Big[ \sum_{\pi_j^i \in \Pi_{r^i_j}^{q^i_j, \bar q^i_j  }  } 
\Tr  \big( O^{\pi_j^i}_0 \sqcup O^{\pi_j^i}_1 \sqcup \cdots \sqcup O^{\pi_j^i}_{r^i_j} \big) 
\Big]  \Big\}\label{faafullsum}
\ee
where the indices of the previous \eqref{faasum} are simply all decomposed into indices for each loop $j=1, 2$
of each loop vertex $i = 1 ,\cdots , n $.

Exactly as in \cite{KRS}, we simply glue the $\sqcup$ symbols of \eqref{faafullsum} into $n+1$ traces $\Tr$.
This is the fundamental common feature of the LVE-LVR. 
Each trace acts on the product of all corners operators $O^c$ cyclically ordered in the way obtained by  turning around the connected components $\bar T$. Hence 
we obtain, with hopefully transparent notations,
\bea 
\partial_T^M   \cR^{k}_n  \Big|_{x_{ij} = x_{ij}^T (w)} &=&
\prod_{i=1}^n \prod_{j= 1}^2
\sum_{\pi_j^i \in \Pi_{r^i_j}^{q^i_j, \bar q^i_j  }  } 
\big[\Tr  \prod_{c  \; \circlearrowleft \; \bar T}O^{c}(M^{\dagger}M)^{\eta_{c}} \big]. 
\label{faafullsumcycl}
\eea

\subsection{Proof of the part concerning  R}
\label{bound}

For the remainder part $ {\cal R}^{\mathcal K}_{n}(\lambda,N,J)$, it is a sum over the \emph{LRV tree amplitudes} 
with $2{\mathcal K}$ cilia and $e(T)\geq n+2$, and therefore we can use \eqref{faafullsumcycl}. Hence we write
\bea
{\cal R}^{\mathcal K}_{n}(\lambda,N,J)\!&=&\! \hskip-.5cm
\sum_{\substack{\substack{T\text{ LVR tree }\\ {{\text{ with }} 2{\mathcal K} \text{ cilia}},}\\ e(T)\geq n+2}}     
\int \{dw_{T} dt du dv\} \Phi_n F^{\mathcal K}_T (\lambda,N,J, v) \hskip-.2cm \prod_{f\in b(G)} \hskip-.1cm\Tr[(JJ^\dagger)^{c(f)}] ,
\nonumber \\
F^{\mathcal K}_T (\lambda,N, v)  &=&\hskip-.2cm \int d\mu_{C\{x_{ij}^\cT\}}(M)\prod_{i=1}^n \prod_{j= 1}^2 
\sum_{\pi_j^i \in \Pi_{r^i_j}^{q^i_j, \bar q^i_j  }  } 
\big[\Tr  \prod_{c  \; \circlearrowleft \; \bar T}O^{c}(M^{\dagger}M)^{\eta_{c}} \big].
\eea

We now bound the functional integral. 
Since there are exactly $n+1$ traces, 
the factors $N$ \emph{exactly cancel}, all operator norms now 
commute and taking into account  \eqref{faacomb} we are left with
 \bee   \vert  F^{\mathcal K}_T (\lambda,N, v)   \vert  \le K^n
\int d\mu_{C\{x_{ij}^\cT\}}(M) \prod_{i=1}^n r_i ! 
\Bigl[ \prod_{c\in \bar T}  \Vert O^c  (M^{\dagger}M)^{\eta_{c}}\Vert \Big]_{x_{ij} =\frac{x_{ij}^T (w)+ x_{ji}^T(w)}{2}}. \label{funcbou1}
 \ee
 where, like in \cite{KRS}, $K$ is a constant for $1 \le {\mathcal K} \le {\mathcal K}_{max}$.
 
Exactly as in \cite{KRS}, using that $\sup\{  \Vert M \Vert , \Vert M^\dagger \Vert  \}\le \Vert M^\dagger M \Vert^{1/2}$, 
it is easy to now bound, for $v$'s on these keyhole contours, the norm of resolvent factors such as $\Vert \frac{1}{v^i_j - X^i} \Vert$ by
a constant times $ (1 + \vert v^i_j  \vert)^{-1}$ and the norm of resolvent factors such as $\Vert \frac{1}{v^i_j - X^i} M^i  \Vert $ or $\Vert {M^i}^\dagger \frac{1}{v^i_j - X^i} \Vert$ by a constant times $(1 + \vert v^i_j  \vert)^{-1/2}$.
 Plugging into
\eqref{funcbou1} we can use again \eqref{faacomb} to prove that we get exactly a decay factor $ (1 + \vert v^i_j  \vert)^{-(1+ r^i_j/2)} $
for each of the $2n$ loops. The corresponding bound being \emph{uniform} in all $\pi, \{w\}, \{M\}$, and since the integral
$\int d\mu_{C\{x^T\}}$ is \emph{normalized}, we get
\bee   \vert  F^{\mathcal K}_T (\lambda,N, v)  \vert  \le K^n\; 
\prod_{i=1}^n \Big\{ r_i ! 
\prod_{j= 1}^2  (1 + \vert v^i_j  \vert)^{-(1+ r^i_j/2)} \Big\}, \label{funcbou2}
\ee
where, again, $K$ is a constant for $1 \le {\mathcal K} \le {\mathcal K}_{max}$.
Recall that with our notations, $r_i =  r^i_1 +  r^i_2$.
Since all integrals with respect to $w$ are  \emph{normalized}, i.e.
\bee \int dw_{T}   \phi(w_{T}) =  \prod_{(i, j) \in T} \int_0^1 dw_{ij} \phi (\{ w_{ij}\}) \le \Vert \phi ( \{ w_{ij}\} )\Vert,
\ee
we have simply to bound
\bea
\int \{dw_{T} dt du dv\} \Phi_n \le \Vert \int \{dt du dv\} \Phi_n \Vert .
\eea
Now we bound the contour integral bound $\int \{ dt du dv\} \Phi_n$ exactly like in \cite{KRS}.
Finally since each vertex has at least one contour operator, the number of $\vert \lambda \vert^{\frac{1}{4p^2}}$ factors in the bound is at least $n$. 
Taking into account that the number of (labeled) trees 
$T$ is bounded by $K^n n!$ for some constant $K $ (for, again, $1 \le {\mathcal K} \le {\mathcal K}_{max}$),
we arrive at 
\bea\label{mainbou}
{\cal R}^{\mathcal K}_{n}(\lambda,N,J)&=&
N^{{\mathcal K}}  \Vert J^{\dagger}J\Vert^{\mathcal K} \sum_{n = {\mathcal K}}^\infty K^n
\vert \lambda\vert^{n+2+\frac{n}{4p^2}}  \quad\\
&\le&  K^{\mathcal K} N^{{\mathcal K}}  \Vert J^{\dagger}J\Vert^{\mathcal K},
\eea
uniformly in $N\in{\mathbb N}^*$, J such that $\|J^{\dagger}J\|<\epsilon_{\lambda}$.

\subsection{Conclusion}
So where is, should we say, the crux of Theorem  \ref{th1}? It is in the  ``Borel-LeRoy part"  of the perturbative expansion!

For the  perturbative expansion,  defined in \eqref{treeexpansion01},
\bea&&\hskip-1.4cm
\sum_{\substack{\substack{G\text{ labeled ribbon graph}\\\text{with ${\mathcal K}$ cilia},}\\ e(G)\leq n}}\hskip-.3cm
\frac{(-\lambda)^{e(G)}N^{\chi(G)}}{v(G)!}\hskip-.1cm
\prod_{f\in b(G)} \hskip-.1cm\Tr (J^{\eta_{c}})  + \hskip-.2cm\sum_{\substack{\substack{(G,T)\text{ LVR graph }\\ \text{  with ${\mathcal K}$ cilia,}}\\ e(G)= n+1}}
\hskip-.3cm {\cal A}^k_{(G,T)}(\lambda,N,J) ,
\eea
the analytic part of the proof of Theorem  \ref{th1} is obvious, since this part is a \emph{polynomial} with respect to $\lambda$;
but this polynomial is of order 
\bee
e(G)\le (p-1)e(T)\le (p-1)(n+1),
\quad 
 \ee
 so, when we arrived  at \eqref{BorLeRSum0}, we inevitably transform a factor $(qn)!$
 into a factor $[(p-1)n]!$.
\medskip

\section{Appendix: Borel-LeRoy-Nevanlinna-Sokal theorem}

We recall the following theorem \cite{Nev,Sokal,CaGrMa}. 
\begin{figure}[htb]
\[
\begin{array}{cc}
\includegraphics[width=4cm]{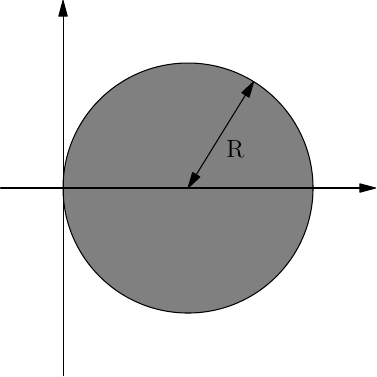}&
\includegraphics[width=6.7cm]{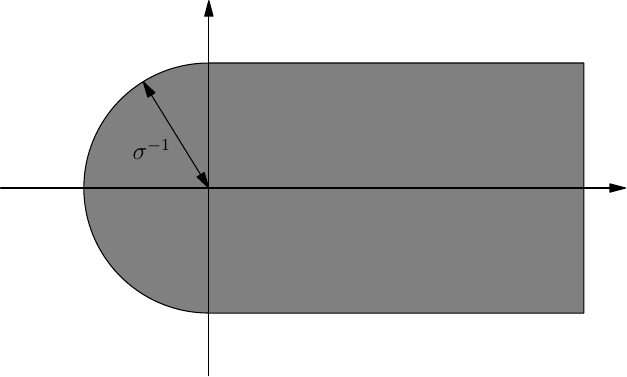}\\
{\cal D}_{R}&\Sigma_{\sigma}
\end{array}
\]
\caption{Domain of analyticity of $F$ and of its Borel transform for $q=1$.}
\label{Borel:pic}
\end{figure}

\begin{theorem}
\label{BorLeRSum}
Let $q\in {\mathbb N^*}$. Let  $F_{\omega}(z)$ be a family of analytic functions on the domain 
\bee
D_R = \{z:\Re z^{-\frac{1}{q}} >(2R)^{-1}\} = \{z:  | z | < (2R)^q \cos^q (\frac{\arg z}{q}) \} 
\label{BorLeRSumbis0}
\ee
depending on some parameter $\omega\in\Omega$, and such that, for some $\sigma \in {\mathbb R}_+$,
\bee
|R_n(z)| =  \big| F_{\omega}(z)-\sum_{m=0}^{n}a_{m}(\omega)z^{m} \big|\le 
\sigma^n
(qn) ! \, | z |^{n+1} \label{BorLeRSum0}
\ee
uniformly in $D_R$ and $\omega\in\Omega$. Then the formal expansion  
\bee\sum_{n=0}^\infty s^{qn} \frac{a_n(\omega)}{(qn)!}\label{BorLeRSumbis1}
\ee
 is convergent for small $s$ and determines a function 
$ B_\omega(s^q )$ analytic in 
\bee
\Sigma_\sigma =\{s :{\text dist}(s ,{\mathbb R}_+ ) < \sigma^{-1} \} 
\label{BorLeRSum1}
\ee 
and such that 
\bee
| B_\omega(s^q )| \le  B \exp\big(\frac{| s |}{R}\big)
\label{BorLeRSum2}
\ee
uniformly for $\Sigma_{\sigma}$ (in \eqref{BorLeRSum2} $B$ is a constant, that is, it is independent of $\omega$). Moreover, setting $t = s^q $,
\bee
 F_{\omega}(z)= \frac{1}{qz}\int_0^\infty B_{\omega}(t)\;
 \Big(\frac{t}{z}\Big)^{\frac{1}{q}-1}  \exp\bigg(-\Big(\frac{t}{z}\Big)^{\frac{1}{q}} \bigg)\; dt
 \label{BorLeRSum3}
\ee
for all $z \in D_R$. Conversely, if $F_{\omega}(z)$ is given by \eqref{BorLeRSum3}, with the above properties for
$B_\omega(s^q)$, then it satisfies remainder estimates of the type \eqref{BorLeRSum0} uniformly, in any $D_r$
such that $0<r <R$, and in  $\omega\in\Omega$.
\end{theorem}
For Theorem  \ref{th1} in  the core of this article, change $q\to p-1$, $z \to \lambda$,
$\omega\to \big\{ N,J \big\}$, whenever $N\in{\mathbb N}^*, \|J^{\dagger}J\|<\epsilon_{\lambda}$.

\end{document}